\documentclass[CRPHYS,Unicode,manuscript]{cedram}

\title{Computer simulations of the glass transition and glassy materials}

\author{\firstname{Jean-Louis} \lastname{Barrat}}

\address{Universit\'e Grenoble Alpes, CNRS, LIPhy, 38000 Grenoble, France}

\email[J-L. Barrat]{jean-louis.barrat@univ-grenoble-alpes.fr}

\author{\firstname{Ludovic} \lastname{Berthier}
\CDRorcid{0000-0003-2059-702X0000}}

\address{Laboratoire Charles Coulomb (L2C), Université de Montpellier, CNRS, 34095 Montpellier, France}

\address{Yusuf Hamied Department of Chemistry, University of Cambridge, Lensfield Road, Cambridge CB2 1EW, United Kingdom}

\email[L. Berthier]{ludovic.berthier@umontpellier.fr}

\thanks{This work was supported by a grant from the Simons Foundation (Grant No. 454933, L.B.)} 


\begin{abstract} 
We provide an overview of the different types of computational techniques developed over the years to study supercooled liquids, glassy materials and the physics of the glass transition. We organise these numerical strategies into four broad families. For each of them, we describe the general ideas without discussing any technical details. We summarise the type of questions which can be addressed by any given approach and outline the main results which have been obtained. Finally we describe two important directions for future computational studies of glassy systems. 
\end{abstract}

\begin{document}

\maketitle

\section{Introduction}

Computer simulations based on molecular dynamics or Monte Carlo simulations of simple models have played an essential role in developing a microscopic understanding of the structure and dynamics of simple liquids~\cite{FrenkelCiccotti-book1987}.  Even with very limited computing power, it was possible to obtain results that could lead to new insights (e.g., hard sphere crystallisation) or could directly be compared to experiments (e.g., neutron or X-ray scattering).

The reasons for this success are simple. First, liquids do not have any mesoscale microstructure, so that relatively small systems (with periodic boundary conditions to avoid surface effects) can be representative of the bulk. Second, their microscopic dynamics is fast, as can be estimated from a typical diffusion time $D/\sigma^2$, where $\sigma$ is the molecular size and $D$ the diffusion constant, of the order of nanoseconds at most. A simulation covering only a few nanoseconds will correctly sample the representative dynamical processes~\cite{rahman1964correlations,hansen1969phase}.

While the first property may still hold to some extent for glasses, with estimated correlation lengths that remain modest even at the experimental glass transition, the second one breaks down rapidly as the glass transition is approached~\cite{berthier2011theoretical}. Even in modestly supercooled liquids, diffusion times will rapidly grow beyond the nanosecond timescale to become of the order of hundreds of seconds at the experimental glass transition temperature $T_g$. As the simulation procedure is limited by microscopic constraints (e.g. a time step of the order of femto or picoseconds), even today's fastest computers are unable to simulate the relevant dynamical processes over such long timescales. The idea of simulating glassy materials on a computer appears totally impractical at first sight.
 
\begin{figure}
    \includegraphics[width=1.\linewidth]{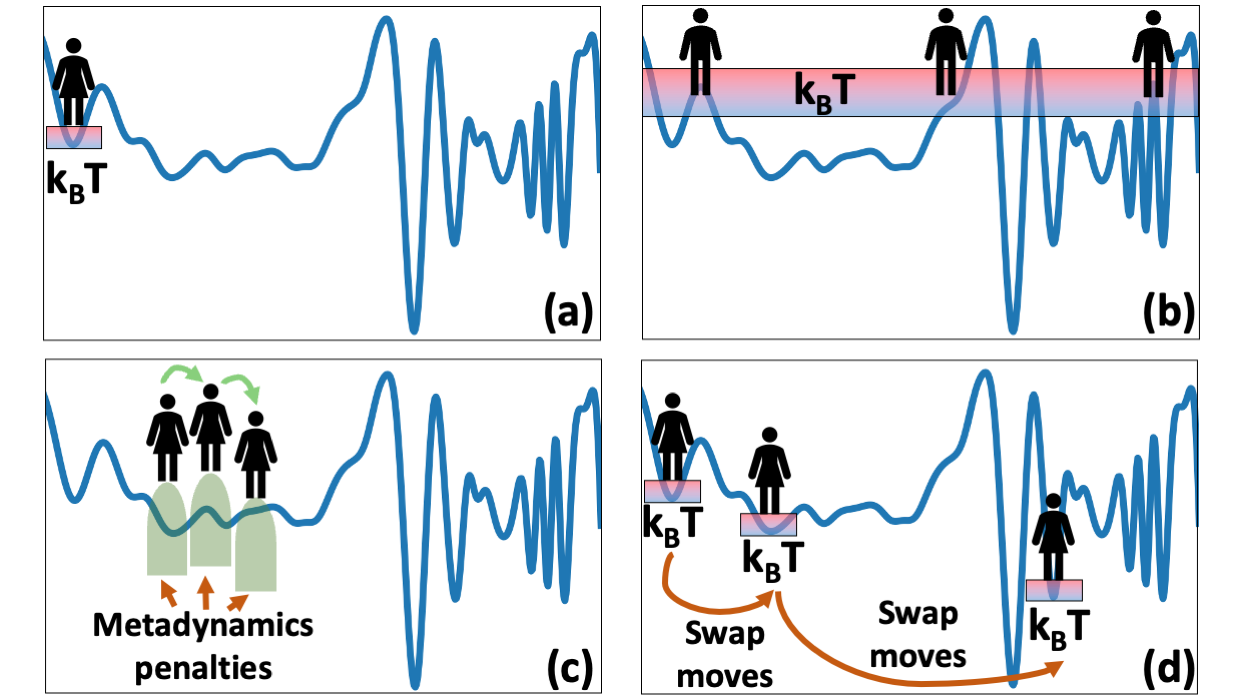}
    \caption{Schematic illustration of possible simulation strategies to study glassy materials. (a) Section 2: the system is quenched into an energy minimum and stays in this minimum, possibly performing small thermal motion. The obtained minimum may be rather atypical, and no improvement is possible.
(b) Section 3: the system is equilibrated at a rather high temperature. Although it starts feeling the presence of the energy landscape, the energy barriers are relatively small and deep energy minima are not accessible to such dynamics.
(c) Section 4: the system is quenched into some energy minimum, and progressively repelled from this minimum by adding energy penalties. New minima are sampled, but their distribution is not controlled by the Boltzmann distribution. 
(d) Section 5: in addition to the thermal motion, unphysical swap Monte Carlo moves allow the system to bypass high energy barriers. An equilibrium ensemble including deep minima is properly sampled, albeit with unphysical dynamics.}
    \label{fig:my_label}
\end{figure}

Several strategies have been proposed over the years to tackle this general problem, which we discuss in this short overview and are illustrated in Fig.~\ref{fig:my_label}. A first one is to create disordered structures ("random packings" or "random networks") by some artificial procedure that may be quite different from the one followed in experiments, the simplest one being to quench instantaneously a liquid configuration to the nearest energy minimum. The hope is then that the structural or vibrational properties of the obtained configuration will bear some similarity to those obtained from the much slower (by many orders of magnitude) cooling protocols realized in experiments.

A second possibility consists in simulating dynamical processes in a moderately supercooled liquid, with relaxation times from a few nanoseconds up to microseconds, and hope that they can be representative of what happens at much lower temperatures and longer timescales. At some point, any simulation will encounter a "computer glass transition" when the relaxation times become larger than the time that can be simulated with the current computer facilities, and again one may hope that this transition is somewhat similar to the laboratory glass transition despite the eight to ten orders of magnitude gap in timescales. 

A third possibility is to develop simulation methodologies that bypass the need for using a time step dictated by the microscopic dynamics, i.e. the so-called "faster than the clock" simulations methods, in which the trajectory of the systems is coarse-grained in time and described as a series of rare events occurring on a slow time scale. When this can be achieved, the microscopic time scale of the simulation then becomes the time scale of the rare event, and very long time scales can in principle be studied at a reduced cost.

Finally, the flexibility of computer simulations can be used to introduce an extended phase space with additional or unphysical degrees of freedom that allow for a much faster exploration of the physical phase space. If this is realized properly, the projection of the corresponding trajectory on the physical phase space will achieve a very efficient sampling, bypassing the slowing down associated with the physical glassy dynamics.

In the following sections, we provide examples in which these different strategies have been employed to explore various aspects of the glassy state or of the glass transition. As we will see from these examples, the questions that can be addressed will be different for the various strategies. For example, a detailed test of statistical physics theories will usually rely on simulations that correspond to a well-controlled, equilibrium statistical ensemble, while a quick exploration of structural properties may be efficiently achieved using less accurate, non-equilibrium methods. Moreover, some strategies will be limited to specific models, and have not yet been extended to describe all types of glasses.   

The field of computer simulation of glasses has expanded enormously in the last decades. A rapid search with keywords "computer glass" returns less than 30 results between 1975 and 1985, and about 1500 during the last decade. Therefore our choice of examples will necessarily be biased by our own interests and expertise. In particular, we will concentrate on relatively simple models that have been used as typical descriptions for  "strong" (network) glasses or "fragile" ones, without any attempt to be exhaustive in terms of materials, phenomenology, and theoretical models.

\section{Random packings and random networks}

Randomness is at the heart of glass physics. A major obstacle to the development of theoretical models of glasses is the aperiodic nature of atomic positions. As a result, a large multiplicity of particle packings may possess equivalent physical properties at the macroscopic scale. Historically, computer simulations have first been employed to generate amorphous structures with macroscopic properties that could reproduce experimental measurements, such as the averaged radial distribution function. In this type of approach, the goal is then to devise a computer algorithm that is able to generate as an output the positions of the atoms which can then be used to measure macroscopic physical observables. The computer then replaces the tedious operation of hand-made model construction used for instance by Bernal~\cite{bernal1960packing,bernal1964bakerian}, Bell and Dean~\cite{bell1966properties} and Polk~\cite{polk1971structural}, by a set of automatic operations performed by the machine. The ultimate goal is of course to obtain a microscopic view of the arrangements of the atoms or molecules composing the amorphous glass structure. 

A simple computational approach was used by Barker and coworkers~\cite{barker1975relaxation} who digitised the ball and stick model manually constructed by Bernal~\cite{bernal1964bakerian}, and then optimised the structure by performing an energy minimisation of the system assuming that balls in fact interact via a Lennard-Jones potential. Such an operation is well-suited for the computer, as this amounts to performing gradient-descent molecular dynamics. The random packing of particles obtained in this manner provides a zero-temperature model of a Lennard-Jones glass, the hope being that it captures the broad features of real glasses made of atoms or very small molecules prepared at finite temperature. 

The idea of a sudden quench to zero temperature starting from a given set of atomic positions remains a central computational tool in glass physics. It was made popular through the landmark papers of Goldstein~\cite{goldstein1969viscous} and Stillinger and Weber~\cite{stillinger1982hidden}, who popularised the concept of inherent states (i.e. energy minima) as being important objects to characterise the properties of the potential energy landscape of glasses~\cite{stillinger2015energy}. In addition, because thermal fluctuations are by definition absent in zero-temperature amorphous packings, it can be hoped that their analysis is easier and can better reveal the intrinsic disorder of glasses rather than the one stemming from thermal fluctuations. Zero-temperature amorphous packings are also commonly used to study the mechanical properties of glasses, as the zero-temperature limit allows again to concentrate on mechanical quantities with no interference and competition with timescales and fluctuations stemming from thermal motion. 

Many glasses of practical interest are not made of spherical atoms but rather of strongly covalently bonded entities, such as the ${\rm SiO}_4$ tetrahedra that form the elementary units of amorphous silica, the main component of window glasses. The conceptual analog of Bernal's random sphere packings for network-forming systems is the idea that such systems can be described as a random network~\cite{wright2013eighty}. This was put forward long ago by Zachariasen~\cite{zachariasen1932atomic}, whose two-dimensional sketch of amorphous silica is maybe the best known molecular representation of a glass in the popular literature. This approach has developed into the idea that network-forming glasses are essentially well-relaxed random networks. Here again, computers have allowed us to shift from hand-made models to specific algorithmic constructions of digital random networks. Over the years, several algorithms to produce random networks have been developed~\cite{wooten1985computer,barkema2000high}, which combine rearrangements of the network topology with energy minimisation or thermally-activated events quantified using particle interactions described by simple potentials such as Keating~\cite{keating1966} or Stillinger-Weber~\cite{stillingerweber1985} interactions. These potentials combine two-body (as in the Lennard-Jones model of liquids) and three-body interactions which favor the formation of open network structures. Over the years, improvement in algorithms and computer power has led to excellent agreement between computer-generated random networks and the measured average structure of real glasses.     

The hypothesis that zero-temperature sphere packings represent an interesting physics problem {\it per se}, with potential connections to the glass transition problem was first put forward by Bernal~\cite{bernal1960packing}. It got revisited and popularised again thirty years ago in the context of the jamming transition~\cite{liu1998jamming,liu2001jamming}. Jamming occurs in the absence of thermal fluctuations when an assembly of repulsive spherical particles undergo a fluid to solid transition as the packing fraction is increased~\cite{o2002random}. Jamming occurs experimentally in non-Brownian particles such as emulsions, large colloidal particles, foams, and granular materials. In the last two decades, analogies and differences between jamming and glass transitions have been well understood, and several original features of amorphous solids near the jamming transition have been described theoretically and studied numerically in great detail~\cite{liu2010jamming,parisi2020theory}. Jammed solids represent a special class disordered solid materials with physical behaviour that differs dramatically from crystals and can be studied numerically using simple models such as soft repulsive particles interacting with harmonic or Hertzian potentials~\cite{durian1995foam}. Although these models do not accurately represent actual molecular or atomic supercooled liquids and glasses, their numerical analysis in the context of equilibrium glassy dynamics~\cite{berthier2009glass}, low-temperature glass properties~\cite{o2003jamming,wyart2005effects}, or rheology~\cite{olsson2007critical,ikeda2012unified} have been very influential.      
\section{Exploring the computer glass transition}

It is a well-known experimental fact~\cite{Brawerbook} that the glass transition takes place at a temperature that depends on the cooling rate. In addition, it is widely accepted that slow relaxations obey, at least to some extent, a "time temperature superposition principle", which states that the slow parts of time (frequency) dependent  correlation  (response) functions behave in a simple way as a function of temperature, i.e. as $f(t/\tau(T))$, or $\chi(\omega \tau(T))$ where $\tau(T)$ is the terminal relaxation time. This approximate principle is commonly used in experimental studies to infer low frequency/low temperature response from high temperature/high frequency data using some approximate extrapolation formula for $\tau(T)$.

Computer studies of the glass transition using standard molecular dynamics simulations use cooling rates in the range of $10^8$~K.s$^{-1}$ or higher. When thermodynamic properties are computed the general phenomenology of the glass transition (e.g. drop in the heat capacity or in the thermal expansion) is recovered, but takes place at much higher temperatures than the laboratory glass transition, at a relaxation time of typically $10^{-8}$~s instead of the $100$~s timescale in experiments. It is now understood that time temperature superposition principle probably does not strictly hold from such short relaxation times to the laboratory glass transition. Still, the system probes time scales that extend over several orders of magnitude beyond the microscopic (vibrational) time scales, and its dynamics already displays nontrivial features such as stretched exponential relaxations and spatial correlations.  Understanding the dynamics in this range and how the system falls out of equilibrium are nontrivial problems that can increase our understanding of experimental observations. 

The first attempt to build a first principle, quantitative theory of the glass transition in simple atomic systems goes back to the development of mode-coupling theories (MCT) by G\"otze and coworkers (reviewed for example in Ref.~\cite{Gotze1992}). Within the framework of this theory,   the long time  dynamics is entirely determined by  static structural properties at the level of pair or triplet correlations, which are easily accessible to computer simulations. At about the same time, pioneering computer simulations~\cite{fox1984molecular,UlloYip1985,mountain1987molecular,Bernu1987,roux1989dynamical,barrat1990diffusion,Barrat1991} started to explore the dynamical properties of supercooled liquids on a time scale of a few nanoseconds, for which glassy features become apparent. Several predictions of MCT concerning for instance the nonergodicity parameter or the amplitude of the so-called $\beta$-relaxation for intermediate times as a function of wavevector, were tested quantitatively. This line of work culminated in the extensive work of Kob and Andersen \cite{Kob1995a,Kob1995b}, who studied in great detail the dynamics of a simple metallic glass model and showed good agreement with the MCT predictions on these short time scales down to a crossover MCT temperature below which deviations become significant. Later studies have confirmed the physical picture that this MCT temperature corresponds to a crossover between a dynamics that mostly explores saddle points of the energy landscape that exhibit a few unstable directions, to a situation dominated by hopping between energy minima~\cite{Cavagna2009}. 

A characteristic feature of a system quenched below its glass transition is the slow time evolution, or aging, of its properties; simple observables such as energy evolve slowly with the waiting time $t_w$ (time elapsed after the quench to low temprature) while two-time response and correlation functions loose their invariance under time translation and become explicit functions of time and waiting time, $C(t,t_w)$~\cite{Struik}. This behaviour can be reproduced, although over a much shorter time scale, in the vicinity of the computer glass transition~\cite{Kob2000}. Typically, one observes a "simple" aging behaviour in which the relaxation time increases linearly with $t_w$. Aging and non-equilibrium response are intrinsically associated with the manner in which the system explores its phase space out of equilibrium. The first theoretical description of this phenomenon was proposed within a framework that bears formal similarities with mode-coupling theories~\cite{kirkpatrick1987p,cugliandolo1993analytical,Cugliandolo1997}. In this description, the out of equilibrium dynamics can be split in two regimes corresponding to two different time scales. On the fast time scale,  the system explores local features of the energy landscape and is in equilibrium with the external thermostat although on a restricted part of its configuration space. On the slow time scale, the dynamics visits the energy landscape out of equilibrium, and effectively behaves as an equilibrium system with a high effective temperature. This "two time scale, two temperature scenario"~\cite{berthier2000two} was extensively tested using numerical simulations both in ageing~\cite{Kob2000} and in driven (sheared) supercooled liquids~\cite{Berthier2002,Ono2002}, with a very good agreement with the theoretical predictions. Subsequent work associated the non-equilibrium effective temperature with the properties of the inherent structures of the potential energy landscape using a thermodynamic approach~\cite{Kob2000}, as well as the fluctuation-response properties of these inherent structures~\cite{Zhang2021}. Overall, this approach can be seen as providing a statistical physics justification of the older empirical concept of a "fictive temperature" commonly used in the materials science community~\cite{Brawerbook}.

Another important feature that was extensively studied in the vicinity of the computer glass transition, in parallel with extensive experimental studies, is the spatially heterogeneous nature of the dynamics~\cite{Ediger2000}. Molecules in different regions of space can move "fast" or "slowly". This feature persists over long time scales and involves a cooperative length scale that increases as the dynamics gets slower. Such dynamical heterogeneities are expected in various theoretical descriptions of the slowing down of liquids~\cite{Toninelli2005,Berthier2011,tarjus2011overview}, and their quantitative study is much easier in simulations than in experiments because numerical trajectories automatically grant access to particle displacements at all simulated times with perfect resolution. Early simulations~\cite{hurley1995kinetic,Kob1997,Donati1998} showed, almost simultaneously with the first quantitative experiments~\cite{Cicerone1995,Bohmer1996}, a clear evidence for the existence of dynamical heterogeneity with a typical size growing as the temperature decreases. Parallel developments of refined experiments and of simulations, using four-point correlation functions and susceptibilities~\cite{lavcevic2003spatially,Berthier2005}, confirmed the interest of quantifying dynamical heterogeneity to test theories of the glass transition. Despite these efforts, current simulation techniques are still unable to probe these heterogeneities in the vicinity of the experimental glass transition, where comparison with theoretical predictions would be very useful. The manner in which dynamical heterogeneity emerges from the local structural disorder remains to be understood~\cite{Shang2019,Berthier2021}. Still, the concepts and methods developed for glasses to quantify heterogeneous dynamics are now widely used in all types of soft, driven, active, and non-equilibrium systems~\cite{Berthier2011}.

While initial studies of the computer glass transition focused on quantities that can also be accessed experimentally (such as dynamical structure factors), subsequent developments also explored a broad range of more complex observables that are for instance useful to more directly test theoretical approaches to glassy slowing down. For instance the study of activity fluctuations and large deviations along equilibrium trajectories~\cite{Hedges2009,speck2012constrained} was compared with predictions arising from dynamical facilitation models~\cite{garrahan2007dynamical}, with good qualitative agreement. In the same spirit, a very recent study~\cite{Chacko2021} used equilibrium simulations slightly below the mode coupling temperature to evidence the role of elastic interactions in mediating facilitation and activity. This could lead to a new view of supercooled  liquids in which an elastoplastic description including thermal activation would provide a consistent description of the dynamics from the mode-coupling temperature down to low temperatures~\cite{Guiselin2022}. 

\section{Accelerated dynamics for glassy systems}

A common view of the dynamics of liquids in the vicinity of the glass transition, or even of glasses in the aging regime, is that it consists of long periods of vibrational motion around energy minima separated by infrequent jumps between distinct minima~\cite{goldstein1969viscous}. Within such a picture, it is tempting to propose approaches in which the periods of vibrational motion are ignored, therefore bypassing the need for a short time step in simulations. The trajectory is then coarse-grained in time as a Markov process visiting successive minima according to prescribed rates.

This strategy is at the heart of the so-called kinetic Monte Carlo approach, which was originally formulated as a Monte Carlo algorithm for systems with a discrete set of microscopic configurations~\cite{Bortz1975}. In this simplified case, the list of all possible transitions starting from a given microscopic states can often be constructed explicitely, and transition rates assigned to each one of them. A rejection-free Monte Carlo trajectory can then be constructed which is statistically exact and avoids the unnecessary unsuccessful trial moves (the equivalent of fast vibrations).

Extending this approach to systems with continuous degrees of freedom is challenging, but can be achieved whenever the set of possible microstates is easily categorized using an underlying discrete representation, as would be the case for example in surface diffusion on a crystalline surface, or for structural defects in crystalline solids~\cite{Chatterjee2007}. A potential difficulty remains in assigning the rates, but many methods, starting with transition state theory, can be used for this purpose.

Glassy and amorphous systems pose a much greater difficulty: the different possible minima are not easily mapped onto a discrete representation, and the possible transitions from one minimum cannot be easily listed and identified, except perhaps for exceedingly small systems~\cite{Angelani1998}, because the complexity of the landscape grows exponentially with system size~\cite{stillinger1999exponential}. As a result, an empirical search for possible transitions must be attempted in order to approximate the dynamics, with no guarantee that the resulting trajectory will be a good approximation of the actual one, or sample a well-defined statistical distribution.

Two main approaches have been used to perform this task. One is the activated relaxation technique proposed by Mousseau and Barkema~\cite{Barkema1996}, in which the transitions are obtained by moving one or a few particles away from their equilibrium positions along a low-energy direction until a saddle point is found (activation stage). Relaxation from the saddle point to a new minimum  follows. This method, by construction, selects events that are local in space. It is therefore likely to be somewhat more realistic for network systems like  amorphous silicon (the case for which it has been used most)~\cite{Trochet2017}.

A more collective and agnostic approach is metadynamics, which globally repels the system by adding a smooth energy penalty function to the current state. The energy is then minimized until a new minimum is found that is not directly affected by the penalty. The method is then repeated, leading effectively to a random walk in the potential energy landscape with barriers between successive minima that can be characterized. In principle, the memory of all previous penalties should be kept during the simulation, but the complexity of the landscape is such that keeping a finite memory is sufficient for an efficient sampling of phase space. For appropriate parameter choices, low-lying energy minima can be reached~\cite{Thirumalaiswamy2022}, although it is unclear how deep they really are compared to slow cooling protocols. With extra assumptions, such as assuming a Markovianity of the generated random walk, this method was used to estimate time correlation functions and transport coefficients in a range inaccessible to molecular dynamics simulations~\cite{Kushima2009}. Although the quantitative results are reasonable, the many approximations involved and the lack of a clear estimate of their consequences prevent direct tests of fundamental theories of the glass transition.

\section{Bypassing the physical dynamics to build ultrastable computer glasses}

A considerable advantage offered by computer simulations is the possibility to manipulate the geometry, control parameters, spatial dimensions, particle interactions and particle motion in ways that defy physical laws in order to achieve a specific goal more efficiently. In this section, we discuss how introducing additional degrees of freedom to the system may lead to a more efficient exploration of the configuration space. 

A paradigmatic example of such a strategy is parallel tempering, also known as replica exchange~\cite{marinari1992simulated,hukushima1996exchange}. In this technique the simulation introduces several copies of the same system that are run in parallel at different temperatures. Monte Carlo moves are then added which attempt the exchange of pairs of configurations evolving at nearby temperatures. One can equivalently consider that each system evolves with a temperature that is no longer a constant, but has become a stochastic variable which act as an additional degree of freedom. Intuitively, one can hope that a configuration blocked within a very stable part of the energy landscape can escape when temperature is increased in the course of the simulation. Parallel tempering has been employed in countless applications~\cite{earl2005parallel}, in particular for systems with a rugged landscape such as protein folding or systems with quenched disorder like spin glasses. The method has also been applied to bulk supercooled liquids, with rather modest success~\cite{yamamoto2000replica,flenner2006hybrid}. A first well-known drawback of the method is that it scales unfavourably with system size and only relatively small systems can be simulated. Second, although the method works well, in the sense that the different replicas appear to travel well across the range of temperatures set by the simulation, achieving equilibration remains quite slow and the reported speedup are modest, a factor of order 10-100 at most. An additional  drawback is that particle motion can no longer be used as a test for proper thermalisation, which makes stringent equilibration tests complicated. The relative lack of success compared to systems such as spin glasses may have an entropic origin. If the complexity of supercooled liquids changes strongly with temperature, an exponential number of novel states are discovered when the temperature is increased, which prevents the fast exploration of a very deep state. In other words, the free energy landscape of supercooled liquids is presumably too chaotic to be efficiently explored using parallel tempering. 

Intuitively, adding more degrees of freedom should accelerate the dynamics. It can therefore come as a surprise that removing (or, "pinning") some degrees of freedom can help preparing highly stable glassy configurations and, more generally, can be considered as something worth doing at all. In the last decade or so, computational studies have nevertheless explored the idea of pinning the positions of a well-chosen set of particles. Why is this a good idea? The first argument is rather generic. If the positions of a subset of the particles are pinned within a bulk configuration taken from the Boltzmann distribution, then the configuration right after pinning is also an equilibrium configuration for the pinned system. Since the pinned system is more constrained than the original one, achieving equilibration in the presence of the pinned particles from scratch would have been very difficult. As a result, extremely stable glassy configurations of the pinned system can be prepared at absolutely zero computational cost. The demonstration that very stable glassy configurations are produced in this manner was provided in Ref.~\cite{hocky2014equilibrium}. More generally, random pinning can be used to  decrease considerably the number of available states, possibly to the point that an ideal glass transition to a glass phase with vanishing configurational entropy can be induced~\cite{kob2013probing,cammarota2012ideal}. 

A second argument demonstrating the usefulness of pinning is that the geometry of the selected set of pinned particles can be varied at will~\cite{berthier2012static}, which opens new ways to tackle an array of different questions. For instance, if an infinite layer of particles is pinned, this creates an amorphous wall that influences the static and dynamic properties of the fluid in its vicinity. Information about important correlation lengthscales can then be  deduced easily~\cite{scheidler2002growing,kob2012non}, as the amorphous wall does not induce any structural perturbation. Another geometry which has been extensively studied is when the position of all particles outside a finite size cavity are frozen from a bulk equilibrated system~\cite{cavagna2007mosaic,biroli2008thermodynamic}. By varying the size of the cavity one can analyse the response of the confined fluid to the frozen boundaries. Theoretical arguments suggest that an important correlation lengthscale called the point-to-set correlation lengthscale can be inferred~\cite{montanari2006rigorous}, and is directly related to the configurational entropy of the bulk system~\cite{bouchaud2004adam}. 

In the above pinning argument, we suggested that preparing a single equilibrium configuration of a liquid deep in the landscape is very easy. However, producing additional equilibrium configurations in the presence of the new constraints remains extremely difficult, as the physical dynamics permitting such exploration is slowed down by the set of pinned particles. In order to perform actual measurements of the physical properties of the pinned system, one faces again the slow dynamics problem. In these highly constrained situations, it turns out that parallel tempering provides considerable help~\cite{kob2013probing,berthier2016efficient,yaida2016point}, at odds with bulk simulations. The physical explanation is presumably that in situations like the frozen cavity or the random pinning close to the putative ideal glass transition, the configurational entropy of the system is fairly small (or even vanishing~\cite{ozawa2018ideal}), so that parallel tempering is no longer plagued by the entropic problem that makes it inefficient in the bulk. 

A classical approach to accelerate convergence to equilibrium and exploration of configuration in computational studies of statistical physics problem is to introduce physically-motivated configurational changes in the course of Monte Carlo simulations. Although originally introduced~\cite{metropolis1953equation} as a tool to efficiently sample the configuration space according to the Bolztmann distribution with no attempt to describe the physical dynamics, Monte Carlo simulations of supercooled liquids based on basic translational moves of the particles provide an efficient method to study both static and dynamics properties of supercooled liquids~\cite{berthier2007monte}. Such Monte Carlo simulations are more efficient than Brownian simulations, and essentially equivalent to molecular dynamics based on Newtonian dynamics. They represent a valid alternative to molecular dynamics, which can be useful for specific applications~\cite{berthier2007efficient,berthier2007revisiting}. 

In addition, it is possible to introduce unphysical particle moves in Monte Carlo simulations in order to accelerate the exploration of the configuration space. Although information about the physical dynamics is lost, this becomes useful whenever the additional moves provide a significant speedup and take the system across barriers that would take very long to cross via the physical dynamics. Such approaches have a long history in statistical physics, and algorithms such as the Wolff cluster Monte Carlo algorithm~\cite{wolff1989collective} have become classic textbook examples. Usually, a good knowledge of the physical problem at hand is needed to propose Monte Carlo moves that significantly weaken the source of the slowing down. For supercooled liquids, it is still unclear a priori what type of collective Monte Carlo move can provide a significant gain. As a result, different sorts of Monte Carlo algorithms have been tested, such as displacing chains of particles~\cite{isobe2016applicability} or performing pivot moves around selected particles~\cite{santen2000absence}, but they both provide modest speedup. 

In 2002, Grigera and Parisi introduced swap particle moves in addition to translational moves to simulate a binary mixture of soft spheres of different sizes~\cite{grigera2001fast}. In this approach a pair of distinct particles is randomly selected and an exchange of their positions is attempted, which is accepted with an acceptance rate that satisfies detailed balance. Equivalently, one can think of a swap move as an exchange of diameters between two particles. In this view, the diameter of each particle becomes a stochastic variable, just as its position. The system is then augmented by one degree of freedom per particle. Later work suggested that the obtained speedup for this algorithm was a constant factor of about two orders of magnitude~\cite{fernandez2006critical}. For this and other systems, swap Monte Carlo led to easy crystallisation of the system~\cite{brumer2004numerical} which slowed down its use and development. It was more recently realised that by optimising both the size distribution of the glass-forming model and the parameters of the swap Monte Carlo algorithm itself, a large, temperature-dependent speedup could be obtained in several representative models of glasses~\cite{berthier2016equilibrium,ninarello2017models}. The speedup is so large that it is difficult to quantify, but can be much larger than 10 orders of magnitude in three-dimensional systems. Physically, it is the coupling between positional and diameter degrees of freedom which holds the key to the observed acceleration of the dynamics~\cite{ninarello2017models}.

The fast equilibration provided by the swap Monte Carlo algorithm means that it becomes possible to prepare equilibrium configurations of supercooled liquids at temperatures where the physical dynamics would be completely arrested on the computer timescale: the computer glass transition is thus fully bypassed, and deeply supercooled states corresponding to temperatures near the experimental glass transition can be prepared in the computer. For some models, temperatures below $T_g$ can even be accessed, thus resulting in ultrastable computer glasses, by analogy with glasses prepared by physical vapor deposition~\cite{swallen2007organic,ediger2017perspective}. Contrary to random pinning, it is possible to prepare a large number of independent stable configurations at a given state point, which then paves the way for the analysis of structural~\cite{wang2019low,coslovich2018local,coslovich2019localization}, thermodynamic~\cite{berthier2017configurational,berthier2019zero}, mechanical~\cite{ozawa2018random,yeh2020glass} and transport~\cite{wang2019sound} properties of supercooled liquids and glasses. Direct tests of theories of the glass transition can be performed in the experimentally relevant temperature regime~\cite{ozawa2019does,guiselin2020random,guiselin2022statistical,guiselin2022static}, as well as exploration of excitations~\cite{berthier2016growing,scalliet2019nature}, and defects~\cite{scalliet2017absence,khomenko2020depletion,nishikawa2022relaxation} characterising low-lying states in the potential energy landscape of glassy systems. The swap Monte Carlo algorithm then inspired novel ways to produce zero-temperature amorphous particle packings with original physical properties~\cite{kapteijns2019fast,varda2022transient}, and was used to understand the peculiar physical properties of ultrastable systems prepared via physical vapor deposition~\cite{berthier2017origin,flenner2019front,berthier2020measure}.

\section{Perspectives}

Nearly sixty years after Bernal's efforts to figure out the disordered atomistic structure of a simple glass, much progress has been done to understand the microscopic properties of supercooled liquids, glasses and the nature of the glass transition separating them. All along this ample research effort, computer simulations have played an important role to address many questions, measure physical quantities, visualise particle motion, and test theoretical ideas across a wide range of models and physical situations. 

This short article demonstrates that multiple approaches have been developed in the field to address a relatively broad diversity of questions regarding the physics of glassy systems. It is therefore not an easy task to quantitatively compare these different methods, as they all have advantages and problems depending on the question asked. Still, we suggest that it would be useful to try and benchmark methods for a specific subset of questions across a range of physical models, to quantitatively establish the relative merits of these methods similar in spirit to what is done for instance in the context of ab-initio methods for condensed matter~\cite{lejaeghere2016reproducibility}.  

Thanks to the various computational strategies described above, the nature of the glass state has been largely elucidated, while its physical properties are much better understood. Yet, research papers are still being written that start with an introductory paragraph that states that the nature of glass is "anything but clear"~\cite{chang2008nature}. What are the main remaining frontiers and can computer simulations help address them? 

An important set of questions remains open regarding the microscopic dynamics of supercooled liquids close to the experimental glass transition. Whereas static and structural properties can now be attacked in this temperature regime, at least for simple models of discrete and continuous mixtures of atoms using the swap Monte Carlo algorithm, it remains impossible to simulate the physical dynamics over the time window that is required to observe structural relaxation over hundreds of seconds. Understanding the nature of the dynamic properties near $T_g$ appears as one of the last frontiers to be attacked numerically. One may hope that combining the rapid equilibration provided by Monte Carlo simulations~\cite{Guiselin2022}, to progress in computational hardware (such as the gain provided by fast GPU simulations over conventional CPU~\cite{bailey2017rumd,coslovich2018dynamic}), to novel methods perhaps aided by machine learning ideas~\cite{schoenholz2016structural,bapst2020unveiling,alkemade2022comparing} and novel physical algorithms will provide a path towards simulating glassy dynamics over long enough timescales.   

A second important frontier for future development is the analysis of glassy materials over a broader range of models, interactions, physical situations than is currently available. Indeed, many of the strategies that we described above have been tailored to relatively simple (not to say simplistic) glass-forming models. While this may be justified from the physical viewpoint by invoking the universality of glass physics~\cite{tarjus2011overview,berthier2011theoretical}, it may be seen as a shortcoming when it comes to the quantitative description of specific physical properties of a specific material. Some of these issues have recently been reviewed in this journal~\cite{liu2022challenges}, describing for instance the emergence of machine learning tools to develop accurate force fields for specific chemical compositions. Also, several of the tools described above, such as the swap Monte Carlo algorithm, have been optimised for simple model systems, but their generalisation to more complex materials, such as molecular or polymer glasses, remains an important goal for future work. 

Finally, there is a wide glassy world outside atomistic and molecular glass-forming materials~\cite{facets2016}. It has become very clear in the last decade that tools and concepts used to understand the glass formation in simple liquids apply more broadly to scores of increasingly complex systems such as colloidal materials~\cite{hunter2012physics}, granular systems~\cite{marty2005subdiffusion,keys2007measurement}, active and biological materials~\cite{berthier2019glassy}. It is certain that simulating these more complex physical situations and materials will shed light on the microscopic mechanisms at play in these systems. There should be a great future in simulating glassy and disordered materials.  

\section*{Acknowledgments}

We thank Cecilia Herrero for her kind help in the realisation of Figure 1.

\section*{Dedication}

We would like to dedicate this article to Jean-Pierre Hansen on the occasion of his 80th birthday.

\bibliographystyle{crunsrt}

\bibliography{samplebib}

\end{document}